\documentclass[sigconf]{acmart}

\usepackage{array}
\usepackage[english]{babel}
\usepackage[utf8]{inputenc}
\usepackage{blindtext}
\usepackage{caption}
\usepackage{subcaption}
\usepackage{multirow}

\usepackage{booktabs}

\usepackage{cleveref}

\usepackage{xspace}
\usepackage{graphicx}
\usepackage{balance}
\usepackage{layouts}

\usepackage[binary-units=true]{siunitx}

\usepackage[absolute,showboxes]{textpos} %
\DeclareSIUnit{\nothing}{\relax}

\let\myRef\ref
\renewcommand\ref{\unskip~\myRef}
\let\myCite\cite
\renewcommand\cite{\unskip~\myCite}

\newcommand{\ie}{i.e., \@}
\newcommand{\eg}{e.g., \@}

\newcommand{\etal}{et al.\xspace}

\newcommand{\openwpm}{OpenWPM\xspace}
\newcommand{\smp}{Subscription Management Platform\xspace}
\newcommand{\smps}{Subscription Management Platforms\xspace}
\newcommand{\cookiewall}{cookiewall\xspace}
\newcommand{\cookiewalls}{cookiewalls\xspace}
\newcommand{\Cookiewall}{Cookiewall\xspace}
\newcommand{\Cookiewalls}{Cookiewalls\xspace}

\newcommand{\parax}[1]{\noindent \textbf{#1}}

\newcommand{\todo}[1]{\textcolor{blue}{\emph{#1}}\xspace}
\renewcommand{\todo}[1]{#1\xspace}

\newcommand{\takeaway}[1]{\emph{\textbf{To summarize: }{#1}}}



\copyrightyear{2023}
\acmYear{2023}
\setcopyright{rightsretained}
\acmConference[IMC '23]{Proceedings of the 2023 ACM Internet Measurement Conference}{October 24--26, 2023}{Montreal, QC, Canada}
\acmBooktitle{Proceedings of the 2023 ACM Internet Measurement Conference (IMC '23), October 24--26, 2023, Montreal, QC, Canada}
\acmDOI{10.1145/3618257.3624846}
\acmISBN{979-8-4007-0382-9/23/10}

\makeatletter
\gdef\@copyrightpermission{
  \begin{minipage}{0.3\columnwidth}
   \href{https://creativecommons.org/licenses/by/4.0/}{\includegraphics[width=0.90\textwidth]{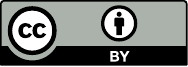}}
  \end{minipage}\hfill
  \begin{minipage}{0.7\columnwidth}
   \href{https://creativecommons.org/licenses/by/4.0/}{This work is licensed under a Creative Commons Attribution International 4.0 License.}
  \end{minipage}
  \vspace{5pt}
}
\makeatother

\begin{document}

\settopmatter{printfolios=true}

\setlength{\TPHorizModule}{\paperwidth}
\setlength{\TPVertModule}{\paperheight}
\TPMargin{5pt}
\begin{textblock}{0.8}(0.1,0.02)
     \noindent
     \footnotesize
	 If you cite this paper, please use the ACM IMC reference:
     Ali Rasaii, Devashish Gosain, and Oliver Gasser. 2023.
     Thou Shalt Not Reject: Analyzing Accept-Or-Pay Cookie Banners on the Web. In \textit{Proceedings of the 23rd ACM Internet Measurement Conference (IMC ’23), October 24–26, 2023, Montréal, Canada}.
\end{textblock}

\title[Thou Shalt Not Reject: Analyzing Accept-Or-Pay Cookie Banners on the Web]{Thou Shalt Not Reject: \\Analyzing Accept-Or-Pay Cookie Banners on the Web}

\author{Ali Rasaii}
    \affiliation{
        \institution{Max Planck Institute for Informatics}
        \country{}
    }
    \email{arasaii@mpi-inf.mpg.de}
\author{Devashish Gosain}
    \affiliation{
        \institution{BITS Pilani Goa}
        \country{}
    }
    \email{devashishg@goa.bits-pilani.ac.in}
\author{Oliver Gasser}
    \affiliation{
        \institution{Max Planck Institute for Informatics}
        \country{}
    }
    \email{oliver.gasser@mpi-inf.mpg.de}

\begin{abstract}

Privacy regulations have led to many websites showing cookie banners to their users.
Usually, cookie banners present the user with the option to ``accept'' or ``reject'' cookies.
Recently, a new form of paywall-like cookie banner has taken hold on the Web, giving users the option to either accept cookies (and consequently user tracking) or buy a paid subscription for a tracking-free website experience.

In this paper, we perform the first completely automated analysis of \cookiewalls, \ie cookie banners acting as a paywall.
We find \cookiewalls on 0.6\% of all queried 45k websites.
Moreover, \cookiewalls are deployed to a large degree on European websites, \eg for Germany we see \cookiewalls on 8.5\% of top 1k websites.
Additionally, websites using \cookiewalls send 6.4 times more third-party cookies and 42 times more tracking cookies to visitors, compared to regular cookie banner websites.
We also uncover two large Subscription Management Platforms used on hundreds of websites, which provide website operators with easy-to-setup \cookiewall solutions.
Finally, we publish tools, data, and code to foster reproducibility and further studies.
\end{abstract}

\maketitle

\section{Introduction}
\label{sec:introduction}

At first glance it may appear, that the vast majority of websites offer their content free of cost. However, many websites have an inherent cost for users by collecting their data and record their personal choices (\eg in the form of cookies), which leads to targeted advertising.
This entire user profiling and targeting nexus is sometimes referred to as ``surveillance capitalism'' \cite{zuboff2019surveillance}. To counter user tracking and safeguard user privacy, privacy laws such as the European Union's GDPR \cite{GDPR} have been enacted.
They mandate that websites take explicit consent from users before storing or sharing their personal data. This led to an increase in cookie banners on websites. These banners notify users about personal data collection policies and provide interaction options like ``accept'' and ``reject'' to the users.

However, recently some websites have switched from showing regular cookie banners to using ``\cookiewalls''.
With \cookiewalls users are given two options---either to provide consent for tracking or to buy a subscription to access a website's content without ads and tracking.
GDPR clearly states that consent to the processing of personal data must be given freely and unconditionally \cite{cookie_banner_taskforce}.
Therefore, the legality of \cookiewalls remains questionable and views of data protection authorities in European countries on the subject differ \cite{morel2022your}.

In this paper, we perform the first completely automated large-scale analysis of the \cookiewall ecosystem to date.
More specifically, the main contributions of this paper are:

\begin{itemize}
    \item \textbf{Large-scale automated measurement study:}
        We perform a large-scale automated measurement study to detect \cookiewalls from eight vantage points on 45k websites.
        We develop a tool to automatically detect \cookiewalls with a precision of 98.2\%, which \todo{we release as open-source \cite{bannerclick} together with our analysis code and data \cite{edmondnew} at \href{https://bannerclick.github.io/}{bannerclick.github.io}} (see \Cref{sec:methodology}).
    \item \textbf{Characterization of \cookiewall landscape:}
        We find \cookiewalls on a total of 280 websites (0.6\%), with some countries such as Germany seeing a 5 times higher prevalence with 2.9\% of reachable websites (see \Cref{sec:measurements:sub:prevalence}).
        We analyze different \cookiewall pricing schemes, finding that around 80\% of websites charge 3 Euro per month or less (see \Cref{sec:measurements:sub:pricing}).
        We investigate if buying a \cookiewall subscription indeed protects from tracking and show that subscribers see no tracking cookies compared to an average of 16 tracking cookies seen by non-subscribers (see \Cref{sec:measurements:sub:cookies}).
        We uncover that the majority of found \cookiewall websites use \smps to facilitate the deployment of \cookiewalls (see \Cref{sec:measurements:sub:smps}).
        We highlight that common ad-blocking solutions are able to block 70\% of \cookiewalls using manually curated filter lists (see \Cref{sec:measurements:sub:ublock}).
    \item \textbf{Discussion of \cookiewall impact:}
        We discuss the impact of the advent of \cookiewalls, reflect on dark patterns to compel users to accept tracking, and reason about the possible future of \cookiewalls (see \Cref{sec:discussion}).
\end{itemize}

\section{Background and Related Work}
\label{sec:background}

In recent years, various data protection laws such as the GDPR \cite{GDPR} in the EU or the CCPA \cite{CCPA} in California have been enacted to regulate the use of Web cookies and other tracking and profiling techniques.
Although sharing similar goals, these laws are implemented in different forms.
GDPR for example mandates that before any storage or exchange of personal information (\eg cookies) the user needs to provide explicit informed consent (\ie opt-in).
CCPA on the other hand states that users must have the option to object to the sharing of their data (\ie opt-out).

Many different scientific works have analyzed the efficacy of these laws \cite{dabrowski2019measuring,sanchez2019can,trevisan2019}, privacy policies \cite{linden2020privacy,degeling2019we,Chen2021,Connor2021}, third parties \cite{sorensen2019before,kretschmer2021cookie}, cookie banners \cite{Santos2021Banners,utz2019informed}, and tracking in the Web \cite{iordanou2018tracing,gonzalez2017cookie, falahrastegar2014rise, englehardt2016online,li2015trackadvisor, schelter2016tracking, Lerner2016, cahn2016empirical,englehardt2015cookies}.

To perform these studies in an automated way, different measurement tools---such as Selenium \cite{selenium}, \openwpm \cite{englehardt2016online,OpenWPM_popular}, FPdetective \cite{acar2013fpdetective}, Chameleon \cite{Chameleon}, and Common Crawl \cite{CommonCrawl}---have been proposed.
In 2021, Jha \etal \cite{jha2021internet} proposed Priv-Accept to accept cookie banners in an automated manner.
In early 2023, Rasaii \etal \cite{rasaii2023exploring} presented BannerClick, a tool which can automatically detect and interact (\ie accept and reject) with cookie banners with an accuracy of 99\% and 96\%, respectively.
Unfortunately, none of the currently available tools is able to automatically detect \cookiewalls.
Therefore, in this work we extend BannerClick to specifically detect these special types of cookie banners, \ie \cookiewalls.

Closest to our research are the works by Papadopoulos \etal \cite{papadopoulos2020keeping} and by Morel \etal \cite{morel2022your}.
In the former paper, the authors investigate paywalls on websites and classify them into soft (limited number of articles can be read before the paywall is shown) and hard paywalls (a subscription is required to access content on a website).
They identify 1.5k websites with some form of paywall-related JavaScript libraries on them.
Contrary to our work, they do not look for \cookiewalls on websites.
In the latter paper, the authors manually annotate and classify \cookiewalls on websites.
They find 13 out of 2.8k websites (0.66\%) showing a \cookiewall to the user.
In comparison to their work, we completely automate the task of detecting \cookiewalls on websites, characterize the prevalent use of \smps among \cookiewalls for the first time, and conduct our study on a much larger set of target websites (more than 45k compared to 2.8k).

\section{Methodology}
\label{sec:methodology}

In this section, we describe the vantage points and target domains for our measurements, detail our \cookiewall detection approach, report on the accuracy of our technique, and discuss the limitations of our approach.

\parax{Vantage Points and Targets:}
We use AWS cloud instances at the following locations as our vantage points (VPs): Frankfurt (Germany), Stockholm (Sweden), Ashburn (US East), San Francisco (US West), Mumbai (India), São Paulo (Brazil), Cape Town (South Africa), and Sydney (Australia). 
\todo{We select these VPs as they include regions with different privacy regulations: GDPR in EU countries (Germany and Sweden), CCPA in California, and LGPD in Brazil. The remaining globally distributed VPs are in countries that have either no or less strict privacy regulations.}

We use Google's Chrome User Experience Report (CrUX) \cite{crux} for target selection, as it was shown to be a more realistic toplist \cite{DBLP:conf/imc/Ruth0WVD22} compared to Alexa \cite{AmazonAl58:online} or Tranco \cite{Tranco}.
We take the union of the country-wise Google CrUX top 10k domains for each VP country, resulting in \num{45222} unique domains reachable in all VPs.

\parax{\Cookiewall Detection Approach:}
To measure the prevalence of \cookiewalls, we use a heavily modified version of the tool BannerClick \cite{rasaii2023exploring}.
BannerClick is built on top of OpenWPM \cite{englehardt2016online} and Selenium \cite{selenium}, and can automatically detect and interact with cookie banners on websites.
We enhance BannerClick by adding support for HTML shadow DOMs \cite{shadowdom} and implement a tailored technique to detect \cookiewalls on websites.

In our tests, we find that multiple websites with \cookiewalls use shadow DOM environments, which can not be directly modified or inspected by browsers or even Selenium \cite{Howcanwe17:online} (\eg it is not possible to look up elements inside shadow DOMs using XPath or CSS selectors).
We work around this limitation by looking for possible elements within the main HTML DOM with the \texttt{shadow\_root} property.
Then we clone and append all child elements within a shadow DOM to the body element of the main document DOM. Thereafter we find the desired button in the cloned DOM and then run the interaction function on the corresponding element in the shadow DOM.
This allows BannerClick to also detect and interact with banners within open and closed shadow DOMs \cite{shadow_open_closed}.

Before detecting \cookiewalls, we first run BannerClick to detect all types of cookie banners.
We then leverage BeautifulSoup \cite{richardson2007beautiful} to search for \cookiewall-specific words and classify banners as \cookiewalls.
As \cookiewalls provide a tracking-free website by paying a subscription fee, we assemble a corpus of \cookiewall-specific words consisting of (1) words related to subscriptions (\ie abo, abonnent, abbonamento, abonne, abonné, ad-free and subscribe) and (2) currency words and symbols\footnote{We use the top 10 global currencies as well as the official currency of our measurement vantage points: EUR, USD, CHF, AUD, GBP, Rs, BRL, CNY, and ZAR.}.
For each currency word or symbol, we check for a possible payment-related combination, \eg \textit{\$3.99}, \textit{3.99\$}, \textit{3.99 \$}, or \textit{3.99 \$}.
If these combinations of currency words or \cookiewall-related words appear in the text of a banner, we classify that banner as a \cookiewall. In total, we find that out of 280 correctly detected \cookiewall websites, 76 make use of a shadow DOM, 132 are embedded in iFrames, and 72 use the main HTML DOM to embed \cookiewalls.
In \Cref{app:screenshots} we show example screenshots for \cookiewalls and cookie banners.
\todo{We release our modified version of BannerClick as open-source software \cite{bannerclick}.}

\parax{Detection Accuracy:}
To measure the accuracy of our \cookiewall detection approach, we randomly select 1000 domains from our target list and manually check their screenshots to find the possible existence of \cookiewalls on the website.
We find that we correctly detect all 6 present \cookiewall websites.
The remaining 994 websites indeed do not show a \cookiewall.
Therefore, for these 1000 random websites we have a precision and recall of 100\%.

Furthermore, we manually check all 285 websites where we detected a \cookiewall to gain confidence in our detection approach.
We find that 280 websites have indeed a \cookiewall, whereas 5 detections are classified as false positives.
This results in a detection precision of 98.2\%.

\parax{Limitations:}
Our study provides valuable insights into the prevalence and characteristics of \cookiewalls. However, it is important to consider certain limitations when interpreting the results:
First,
we use an automated approach with a modified version of the BannerClick tool, achieving a 98.2\% precision rate in detecting \cookiewalls. However, false negatives are still possible, and manual verification may not guarantee complete accuracy for all websites.
Second,
some websites identify web crawlers as bots \cite{KJK22}. Thus when they detect a crawler, they may behave differently---\eg altering the number of cookies or displaying \cookiewalls differently from a regular user. Although OpenWPM has mechanisms to mitigate bot detection, it is impossible to completely circumvent bot detection. Hence, our study may not fully represent the actual website behavior experienced by regular users.
\todo{Third, while our VPs are located in eight different geographical regions across six continents, more VPs in different countries can be added to the study. Thus, future studies can further increase the number of VPs across countries to obtain an even better understanding of \cookiewalls.}  
Finally, our study primarily examines the technical aspects and deployment of \cookiewalls, not user perceptions or behaviors. Understanding user perspectives would require additional research, such as user surveys or studies.

\section{Measurement Results}
\label{sec:measurements}

\begin{table}[!tb]
    \begin{tabular}{lrrrr}
        \toprule
        VP & \Cookiewalls & Toplist & ccTLD & Language \\
        \midrule
        US East & 197 & 0 & 0 & 9 \\
        US West & 199 & 0 & 0 & 9 \\
        Brazil & 196 & 0 & 0 & 0 \\
        Germany & 280 & 259 & 233 & 252 \\
        Sweden & 276 & 15 & 0 & 0 \\
        South Africa & 199 & 0 & 0 & 0 \\
        India & 192 & 0 & 0 & 10 \\
        Australia & 190 & 5 & 0 & 10 \\
        \bottomrule
    \end{tabular}
    \caption{Number of detected \cookiewalls depending on the country of the vantage point, country-specific toplist, TLD associated with that country, and the most commonly spoken language in that country.}
    \label{tab:overview}
\end{table}

In this section, we present results from our \cookiewall measurements, including \cookiewall prevalence across multiple characteristics, subscription pricing, third-party and tracking cookie analyses, a case-study of \smps, and results from experiments to bypass \cookiewalls.

\subsection{\Cookiewall Landscape}
\label{sec:measurements:sub:prevalence}

We use our modified version of BannerClick to run \cookiewall measurements from eight vantage points targeting \num{45222} websites.
In \Cref{tab:overview} we show different characteristics of our measurements and the detected \cookiewall websites.
In total, we find \cookiewalls on 280 unique websites, resulting in an overall \cookiewall rate of 0.6\%, a similar rate as found by previous work on a smaller set of target websites \cite{morel2022your}.
Our vantage points (VPs) in the EU (Germany and Sweden) see around 280 websites with \cookiewalls compared to around 200 for non-EU VPs.
This finding is consistent with the generally higher prevalence of cookie banners in the EU \cite{rasaii2023exploring}.

Next, we analyze different characteristics---\ie country-specific toplists, top-level domains, and language---for \emph{each vantage point} separately.
We find that the Germany-specific CrUX toplist (see \Cref{sec:methodology}) contains by far the most detected \cookiewall websites (259, 2.9\% of reachable top 10k websites), followed by Sweden (15) and Australia (5).
We also find cases where websites on a country-specific toplist show a \cookiewall only when visited from a particular VP\footnote{For example, the website \href{https://pt.climate-data.org/}{pt.climate-data.org} is on the Brazilian country-specific toplist, but only shows a \cookiewall when visited from Germany or Sweden. This particular website is in fact operated by a German person, but provides specific subdomains for different languages, \eg \texttt{pt.} for Portuguese.}.
This shows that \cookiewalls are affecting users differently based on the list of popular websites within their country.

To better understand websites showing \cookiewalls to their visitors, we analyze the website top-level domain (TLD), the website's language, as well as the category the website can be attributed to.
We find that again the vast majority of \cookiewall websites are hosted on Germany's \texttt{.de} country-code TLD (ccTLD), followed by generic TLDs (14 on \texttt{.com}, 14 on \texttt{.net}, 4 on \texttt{.org}), and non-VP ccTLDs (6 on \texttt{.it}, 4 on \texttt{.at}, and 2 on \texttt{.fr}).

\begin{figure}[!tb]
    \includegraphics[width=0.8\columnwidth]{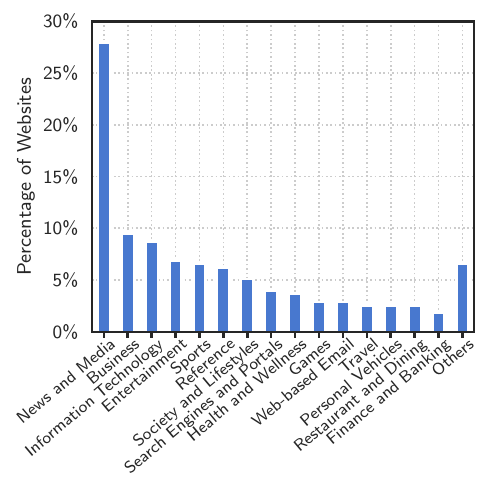}
    \caption{Categories of websites showing \cookiewalls.}
    \label{fig:categories}
\end{figure}

Next, we inspect the language of the \cookiewall websites using CLD3 \cite{cld3} to characterize the main target audience.
Unsurprisingly, the largest part of these websites are in German\footnote{Note that this might also include websites targeted at readers outside Germany, \eg Austria, Switzerland, or other German-speaking audiences.}, followed by English (US, Australia, India), Italian, and Swedish.
To characterize the content of the website, we use FortiGuard's Web filter database \cite{WebFilte22:online} to assign each website to a category.
As shown in \Cref{fig:categories}, more than one-fourth of all \cookiewall websites are categorized as news and media, 9\% fall into the business category, and 7\% are IT-related websites.
This highlights that \cookiewalls---although they are most prominent on news websites---go beyond just news websites and are deployed on a large variety of different website categories.

Additionally, we find that \cookiewalls are more prevalent on popular websites, \ie 1.7\% of country-wise top 1k domains show \cookiewalls compared to 0.6\% for top 10k domains\footnote{Note that the Google CrUX toplist does not contain detailed rank information per website. It rather groups websites into rank buckets, \eg top 1k or top 10k.}.
Interestingly, if we just consider the top 1k reachable websites for Germany, we detect \cookiewalls on more than 8.5\% of websites, almost double the 4.7\% in 2022 \cite{morel2022your}.

\takeaway{\Cookiewalls are most prominent on websites which are popular among users from Germany, where we see them on 2.9\% of top 10k websites and 8.5\% of top 1k websites.
Moreover, \cookiewalls are visible on a wide variety of website categories, with news and media websites making up more than one fourth. In addition, more popular websites are more likely to show \cookiewalls.}

\subsection{Subscription Pricing}
\label{sec:measurements:sub:pricing}

\begin{figure}[!tb]
    \includegraphics[width=0.8\columnwidth]{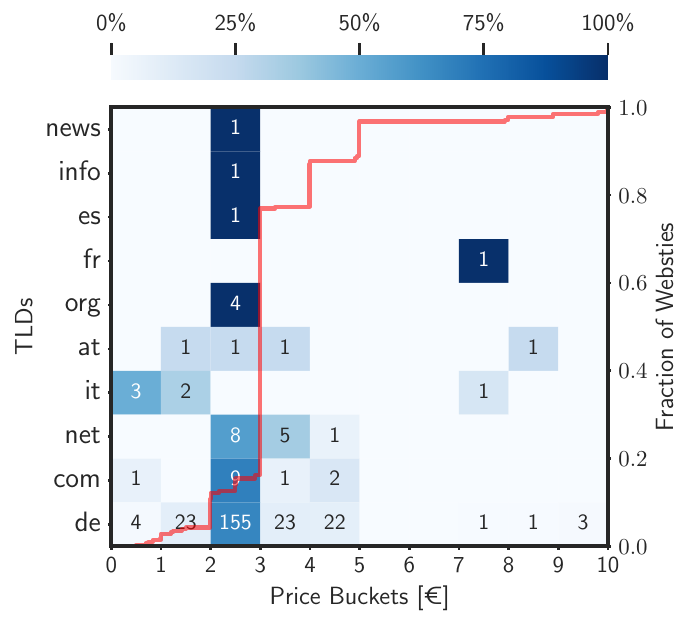}
    \caption{Distribution of monthly subscription price for \cookiewall websites.}
    \label{fig:prices}
\end{figure}

In this section, we analyze the price of \cookiewall subscriptions of all detected 280 websites.
We manually inspect each website to determine the price of a subscription.
Then, we normalize the subscription price by month and convert it to Euro to make different websites comparable.

\begin{figure}[!tb]
    \includegraphics[width=0.8\columnwidth]{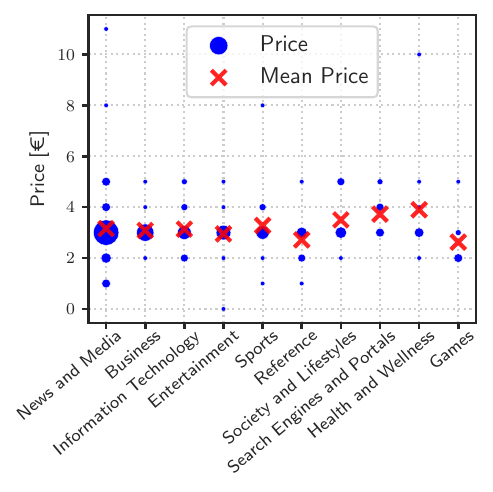}
    \caption{\todo{Correlation between the category of the websites and price of \cookiewall website subscriptions.}}
    \label{fig:correlation-category}
\end{figure}

In \Cref{fig:prices} we show the distribution of the monthly subscription price for \cookiewall websites.
The red line shows an ECDF for the prices of \cookiewalls for all TLDs. We find that around 90\% of \cookiewall websites ask for 4 Euro (approx. 4.33 USD) or less per month, and by far the largest fraction of websites charges 3 Euro (3.25 USD), with the majority of these websites being attributed to a \smp  \todo{in which
subscribers just need to pay once to access all partnered websites (see \Cref{sec:measurements:sub:smps}).}
On the other end, a handful of websites ask for 9 Euro (9.74 USD) or more per month. The heatmap in \Cref{fig:prices} shows the occurrence of each price bucket for each TLD separately. We find that TLDs of websites do not have a substantial impact on the prices, as most websites in different TLDs charge between 2 to 3 Euro per month, except for \texttt{.it} which are on average cheaper. \todo{Furthermore, we explore potential correlations between website categories and subscription prices. In \Cref{fig:correlation-category} the size of the blue data points represents the number of websites falling within each price range, with the red cross showing the mean price per category. We find no obvious relationship between subscription price and website category.}

\takeaway{We find that 90\% of \cookiewall websites charge at most 4 Euro, with some outliers charging upwards of 9 Euro per month. Moreover, we find the prices to be generally similar for different TLDs \todo{and website categories.}}

\subsection{Third-party and Tracking Cookies}
\label{sec:measurements:sub:cookies}

To assess the effect of \cookiewalls on user privacy, we now compare cookies sent by websites with \cookiewalls to websites with ``regular'' banners.
Therefore, we run additional measurements targeting 280 \cookiewall websites and 280 randomly selected websites with regular cookie banners with an accept button.
To account for variations in advertisements and consequently sent cookies, we repeat each measurement five times per website and calculate the average number of cookies per website.
We then compare the number of first-party, third-party, and tracking cookies after accepting \cookiewalls and regular cookie banners.
Similar to previous work \cite{gotze2022measuring,rasaii2023exploring}, we use the justdomains blocklist \cite{justdomains} to classify cookies as tracking cookies.
\todo{If the cookie domain matches one of the domains in the justdomains list, we classify it as a tracking cookie.}
\todo{Note that there exist other techniques to track users that we do not consider in this research, \eg browser fingerprinting \cite{acar2014web}, tracking using first-party cookies \cite{chen2021cookie,munir2023cookiegraph,demir2022towards}, and the use of invisible pixels and click IDs \cite{bekos2023hitchhiker}, as we specifically focus on studying the emergence of \cookiewalls. Thus, in the future, a more nuanced analysis focusing on other tracking techniques can be conducted.}

\begin{figure}[!tb]
    \includegraphics[width=1\columnwidth]{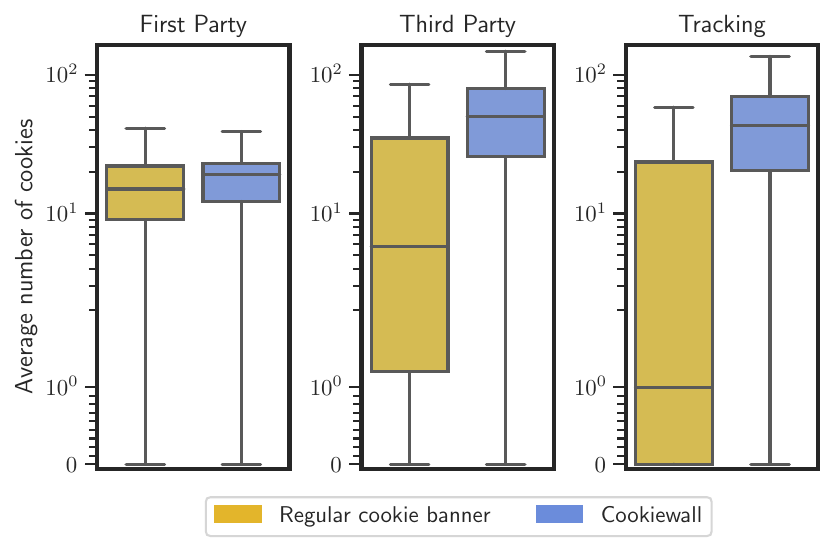}
    \caption{Average number of cookies comparing websites with regular cookie banners to \cookiewall websites.}
    \label{fig:banner_vs_cookiewall}
\end{figure}

\todo{\Cref*{fig:banner_vs_cookiewall} compares the average number of cookies set by websites with regular cookie banners and \cookiewalls.}
In the figure, we see a similar number of first-party cookies among both website sets, with a median of 15 and 19 for regular cookie banner and \cookiewall websites, respectively.
In contrast, third-party cookies exhibit a stark difference between both website sets. We find many more third-party cookies on \cookiewall websites with a median of 50.4, compared to just 6.8 for cookie banner websites.
An even more pronounced discrepancy can be seen for tracking cookies, with \cookiewall websites sending on average 42 times more tracking cookies compared to cookie banner websites (median: 43 vs. 1).
This seems to indicate, that websites with \cookiewalls try to monetize their users more aggressively compared to other websites, either through subscription fees or excessive tracking and advertising.

\takeaway{\Cookiewall websites send 6.4 times more third-party and 42 times more tracking cookies compared to ``regular'' cookie banner websites.
This highlights the focus on monetization efforts of \cookiewall websites.
}

\subsection{Subscription Management Platforms}
\label{sec:measurements:sub:smps}

Similar to Consent Management Platforms (CMPs) for regular cookie banners \cite{hils2020measuring,toth2022dark}, we find two different \smps (SMPs) for \cookiewalls: contentpass \cite{contentpass} and freechoice \cite{freechoice}, which claim to host \cookiewalls for 219 and 167 websites, respectively.\footnote{\todo{We observe an increase in these numbers between May and September 2023 to 270 for contentpass and 184 for freechoice.}}
These two SMPs provide ad-free access to all partner websites for a monthly fee of 2.99 Euro.
Note that only 76 contentpass and 62 freechoice partner websites are in our merged top 10k target list of previous measurements.
We also find evidence of interoperability between CMPs and SMPs, with the CMP consentmanager providing integration support for the SMP contentpass \cite{consentmanager_contentpass}.

In order to contrast the experience of subscribed users and users accepting tracking on SMP websites, we run an additional measurement for all 219 contentpass partner websites.
Thus we create a contentpass account and buy a one-month subscription.
We automate the login behavior on each of these websites and compare the sent cookies to clicking ``accept''.
We again run five repetitions per website and average the number of cookies, in order to take website and advertisements variations into account.

\begin{figure}[!tb]
    \includegraphics[width=1\columnwidth]{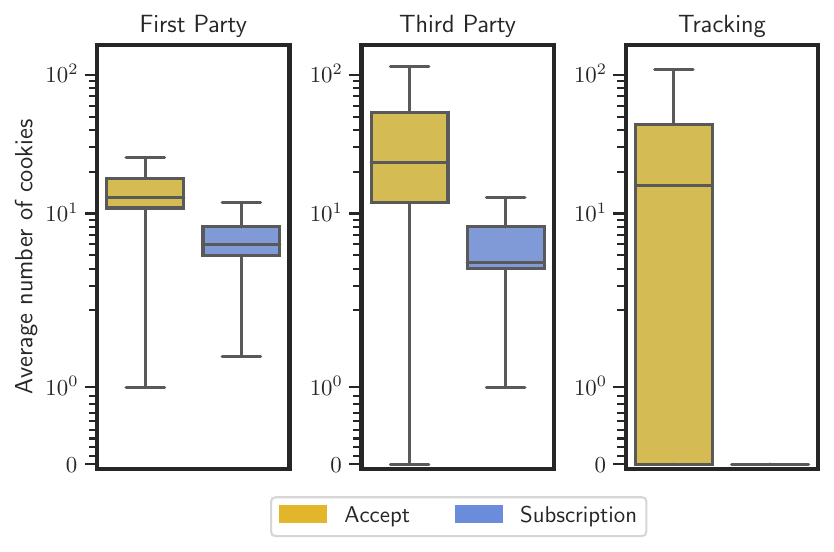}
    \caption{Average number of cookies set by websites with contentpass \cookiewall after accepting or accessing with a subscription.}
    \label{fig:contentpass_accept_signin}
\end{figure}

\Cref{fig:contentpass_accept_signin} shows the distribution of the average number of first-party, third-party, and tracking cookies across all 219 contentpass websites.
We find a lower number of first-party (FP) and third-party (TP) cookies when accessing these websites with a subscription, with a median of 13 vs. 6 FP and 23.2 vs. 4.4 TP cookies for accept and subscription, respectively.
The most apparent difference can be seen for tracking cookies, where we see no tracking cookies with a subscription compared to a median of 16 when accepting the \cookiewall.
Some websites send more than 100 tracking cookies when accessing these websites without a subscription.
This underlines that \cookiewall websites are in fact aggressively tracking users, likely to maximize their income from non-subscribing users via ads and to push them towards buying a subscription.

\takeaway{\smps provide an easy way for website operators to monetize their users by offering them a subscription instead of being tracked and served with ads.
While subscribed users see no tracking cookies, users accepting \cookiewalls of the contentpass SMP see a median of 16 tracking cookies with some extreme cases sending more than one hundred tracking cookies.
}

\subsection{Bypassing \Cookiewalls}
\label{sec:measurements:sub:ublock}

This section delves into the feasibility, implications, and tools available for bypassing \cookiewalls on websites. The forcible accept-or-pay scheme of \cookiewalls might in the eyes of some users justify the act of bypassing it without being concerned about ethical considerations.   
One commonly employed method for bypassing \cookiewalls is the use of ad-blocker browser extensions. Notable examples include ``I don't care about cookies'' \cite{idontcareaboutcookies} ``Ninja Cookie'' \cite{NinjaCoo52:online} and ``uBlock Origin'' \cite{uBlockOr8:online}. In this section, we focus on investigating the effectiveness of uBlock Origin, one of the most popular ad-blocker extensions.

To evaluate its effectiveness, we conduct a measurement on our 280 detected \cookiewall websites. We enable the uBlock Origin extension\footnote{We enable the by default disabled Annoyances filter lists to block \cookiewalls.} and access each of the websites five times. We find that 196 (70\%) websites no longer display \cookiewalls across all iterations, while the remaining websites still exhibit the \cookiewall prompt. 
Note that while browser extensions like uBlock Origin can effectively block resources with domains\footnote{Example of patterns in the block lists which prevent further communication with CMPs to show the banners: \texttt{*cdn.opencmp.net/*}, \texttt{*consentmanager.net/*}, \texttt{*usercentrics.eu/*.}} 
listed in block lists (such as Easylist), they may not perfectly eliminate all types of \cookiewalls. 
Some \cookiewalls may be served locally or use lesser-known third-party domains, which could evade the blocking measures. \todo{Additionally, we manually inspect these 196 websites and find that all of them except two\footnote{\todo{\href{https://hausbau-forum.de/}{hausbau-forum.de} detects uBlock and asks the user for deactivation. \href{https://promipool.de/}{promipool.de} is clickable but not scrollable.}} work normally and do not show any ads.}

\takeaway{Browser extensions like uBlock Origin can effectively block 70\% of \cookiewalls in our measurements.}
\section{Discussion}
\label{sec:discussion}

We now discuss the implications of our findings and present future research directions.

\parax{Paywalls vs. \Cookiewalls:} Existing research \cite{papadopoulos2020keeping} reports on the rise of two types of Internet paywalls---hard and soft. 
With hard paywalls, users cannot access the website without first buying a subscription. With soft paywalls, users can freely view a certain number of articles before they need to buy a paid subscription.
In this paper, we highlight the use of \cookiewalls where users (1) have to pay to \textit{not} opt-in to tracking or, (2) accept using a service with tracking, or (3) cannot access the website's content at all. From a monetary perspective, \cookiewalls are similar to hard paywalls, but overall they adversely impact the clients' privacy. Due to this new ``pay or get tracked'' model, users may be conditioned to accept tracking cookies rather than paying for their privacy.
This can result in privacy laws like GDPR being less effective. Moreover, in the future, websites may charge unreasonably high prices that could further compel users to accept tracking as their default choice. \todo{Although previous researchers \cite{toth2022dark,gray2021dark} also highlight the deployment of manipulative and non-compliant consent pop-ups by different CMPs, they do not consider \cookiewalls. For instance, Toth \etal \cite{toth2022dark} report that CMPs like Quantcast provide configuration interfaces to set up cookie banners and restricted website access, \ie limited (or no access) to website content before interacting with the banners.}

\parax{\Cookiewalls, Website Content, and Tracking Cookies:}
Websites that show \cookiewalls may offer important content to their clients.
We find that many websites showing \cookiewalls are in top 1k of domains.
Thus, users will either provide consent to tracking or pay to avoid it, as they would not want to cease access to the website content.
\Cookiewalls have the potential to create two classes of Web users: those that can afford to not being tracked, and those who need to pay for services with their data.
In the future, user studies can be conducted to estimate the ``monetary value of the content'' on \cookiewall websites.

\begin{figure}[!tb]
    \includegraphics[width=0.75\columnwidth]{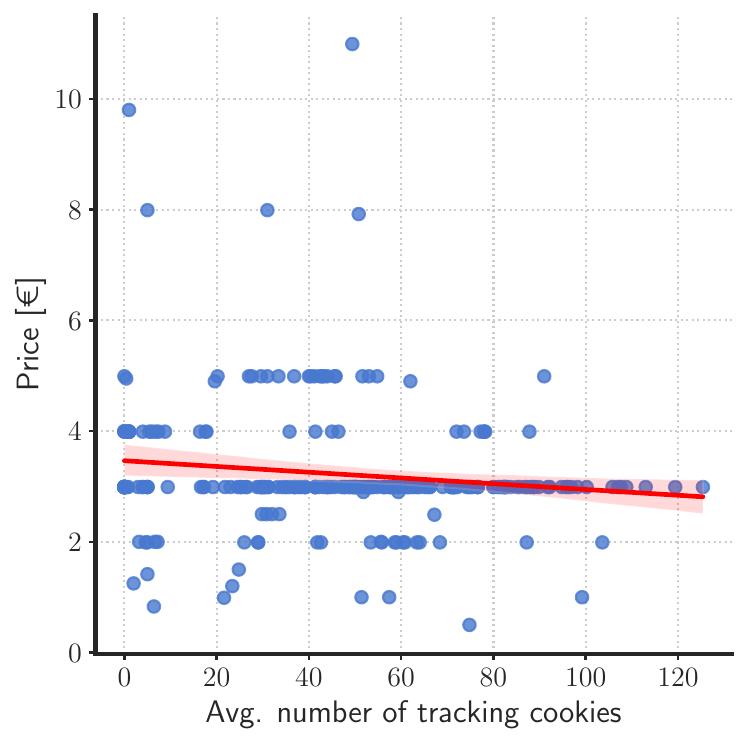}
    \caption{Correlation between the number of tracking cookies and price of \cookiewall website subscriptions.}
    \label{fig:correlation}
\end{figure}

Tracking cookies themselves are used to facilitate ad serving, thus bringing monetary value to the website.
To see if there is a correlation between the number of tracking cookies a website sets for ``accepting'' users and the subscription price, we run an additional small experiment.
\todo{As  shown in \Cref{fig:correlation}, we observe no meaningful linear correlation between the number of tracking cookies set by websites when accepting tracking and the subscription price.}

\parax{Circumventing Banners and \Cookiewalls:}
Interaction with cookie banners could be seen as a nuisance to some users. Thus browsers such as Firefox are working on automatically clicking the reject button (if available) on banners \cite{firefox-reject-banners}. This approach as well as our tool could lay the groundwork to also automatically interact with \cookiewalls in the future.

Presently, we can use ad-block extensions and filter lists to evade most \cookiewalls.
There is, however, a risk that these extensions may also block necessary scripts, potentially disabling useful functionalities, or introducing security threats. Moreover, since ad-blockers run as a script on the client side, they can themselves be a source of privacy leaks.

\parax{Revoking \Cookiewall Acceptance:}  We find that it is not trivial to switch from \cookiewall acceptance
to subscription. If a user has already consented to ``accept'' on some website's \cookiewall, they must delete their cookies and local storage (specific to the website). After deletion, they would see the \cookiewall on a subsequent visit and can change their choice.
Since users will likely not be aware of these necessary additional steps, they might continue to be tracked even though they have subscribed \eg on a different device.

\section{Conclusion}
\label{sec:conclusion}

In this paper, to the best of our knowledge we performed the first automated analysis of the \cookiewall landscape to date.
We developed a tool to automatically detect \cookiewalls with a precision of 98.2\%.
Using this tool we crawled 45k websites and found \cookiewalls on 280 of them.
We investigated different \cookiewall deployment characteristics and uncovered that they are especially deployed among popular websites in Germany (8.5\%).
Moreover, we compared \cookiewalls to regular cookie banners and found websites to be sending 42 times more tracking cookies to \cookiewall website visitors.
Additionally, we uncovered two large \smps which provide website operators with easily deployable \cookiewall solutions.
Finally, we publish our measurement tool to allow for future studies, as well as analysis code and data to foster reproducibility.

\section*{Acknowledgments}

We thank the anonymous reviewers as well as our shepherd Hamed Haddadi for their valuable feedback.
 
\label{body}

\bibliographystyle{ACM-Reference-Format}
\bibliography{paper}

\balance

\appendix
\section{Ethics}

We incorporate proposals by Partridge and Allman \cite{partridge2016ethical} and Kenneally and Dittrich \cite{kenneally2012menlo} and follow best measurement practices \cite{durumeric2013zmap} when running our measurements.
We use dedicated measurement machines, set up informative rDNS names, host a website with information about our measurements, and offer the possibility to be blocklisted from the measurements.
We run \openwpm in a similar way as any regular user when visiting websites with a normal Web browser.
During our measurement period, we did not receive any complaints.

\section{Screenshots}
\label{app:screenshots}

\begin{figure}[!b]
    \includegraphics[width=1\columnwidth]{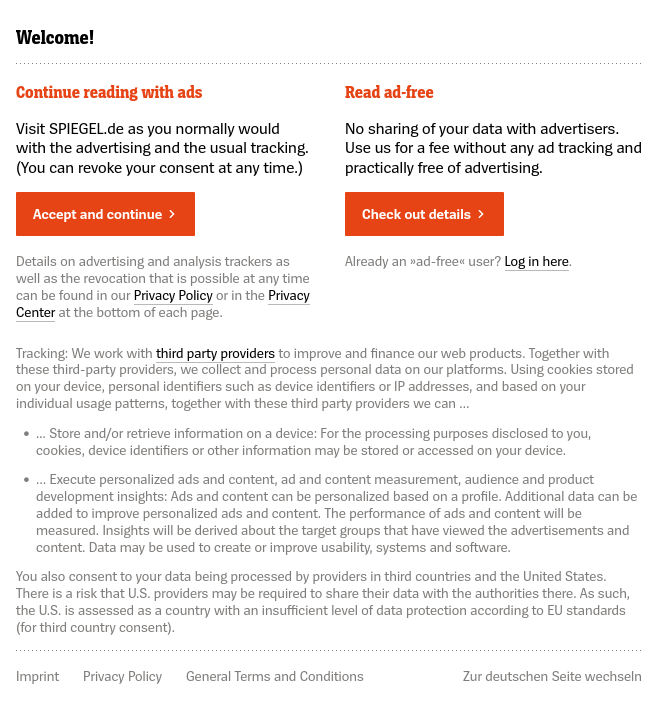}
    \caption{Example of a \cookiewall shown on \href{https://www.spiegel.de}{spiegel.de}.}
    \label{fig:withcookiewall}
\end{figure}

\begin{figure}[!b]
    \includegraphics[width=1\columnwidth]{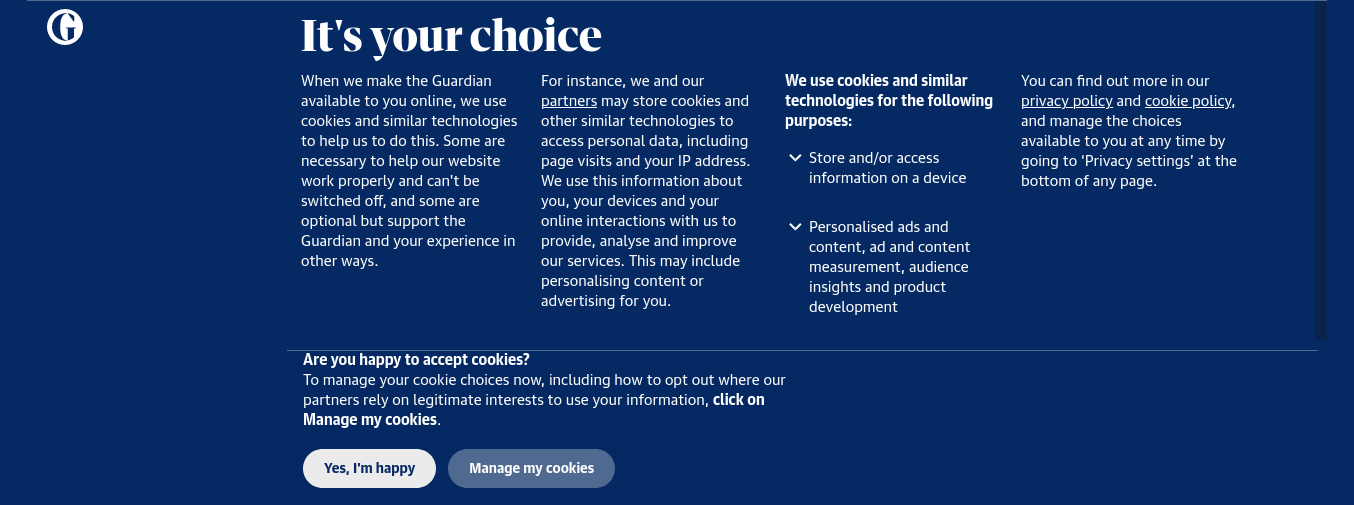}
    \caption{Example of a regular cookie banner shown on \href{https://www.guardian.co.uk}{guardian.co.uk}.}
    \label{fig:withoutcookiewall}
\end{figure}

\Cref{fig:withcookiewall} shows a screenshot of an example \cookiewall on a website, whereas \Cref{fig:withoutcookiewall} shows a screenshot of a regular cookie banner on a website.
Note the presence of a subscription button instead of a ``reject'' or ``options''/``manage my cookies'' button in the \cookiewall.

\label{lastpage}

\end{document}